\documentclass[aps,prl,twocolumn,
showpacs,superscriptaddress,floatfix,nofootinbib]{revtex4}
\usepackage{graphicx}
\usepackage[english]{babel}
\usepackage{amssymb}
\usepackage{amsmath}
\begin{document}

\title{Implications of experimental probes of the RG-flow in quantum Hall systems}
\author{C.A. L\"utken}
\affiliation{Theory Group, Department of Physics, University of Oslo}
\author{ G.G. Ross}
\affiliation{Rudolf Peierls Centre for Theoretical Physics, Department of Physics,
University of Oxford}
\date{\today}

\begin{abstract}
We review the implications of the scaling data for the emergent symmetry
of the quantum Hall system. The location of the fixed points in the
conductivity plane is consistent with the global, non-Abelian discrete
symmetry $\Gamma _{0}(2)$, and the renormalisation group (RG) flow-lines agrees
closely with that found if the symmetry acts anti-holomorphically. We extend
the analysis to consider the rate of the RG flow. 
For a specific model in
which the $\Gamma _{0}(2)$ symmetry acts anti-holomorphically the scaling
close to the fixed points gives a critical delocalisation exponent 
$\nu = 2.38\pm 0.02$, 
in excellent agreement with direct measurements and with
numerical simulations. Both the predicted flow-lines and the flow
rate also agree with the experimental measurements far away from the critical
points, suggesting an  emergent topological structure capable of
stabilising the symmetry predictions. We hope that this agreement will
stimulate further experimental study capable of conclusively testing the
symmetry and exploring its associated dynamics.
\end{abstract}
\maketitle

\begin{figure}[t]
\begin{center}
\includegraphics[scale = .54]{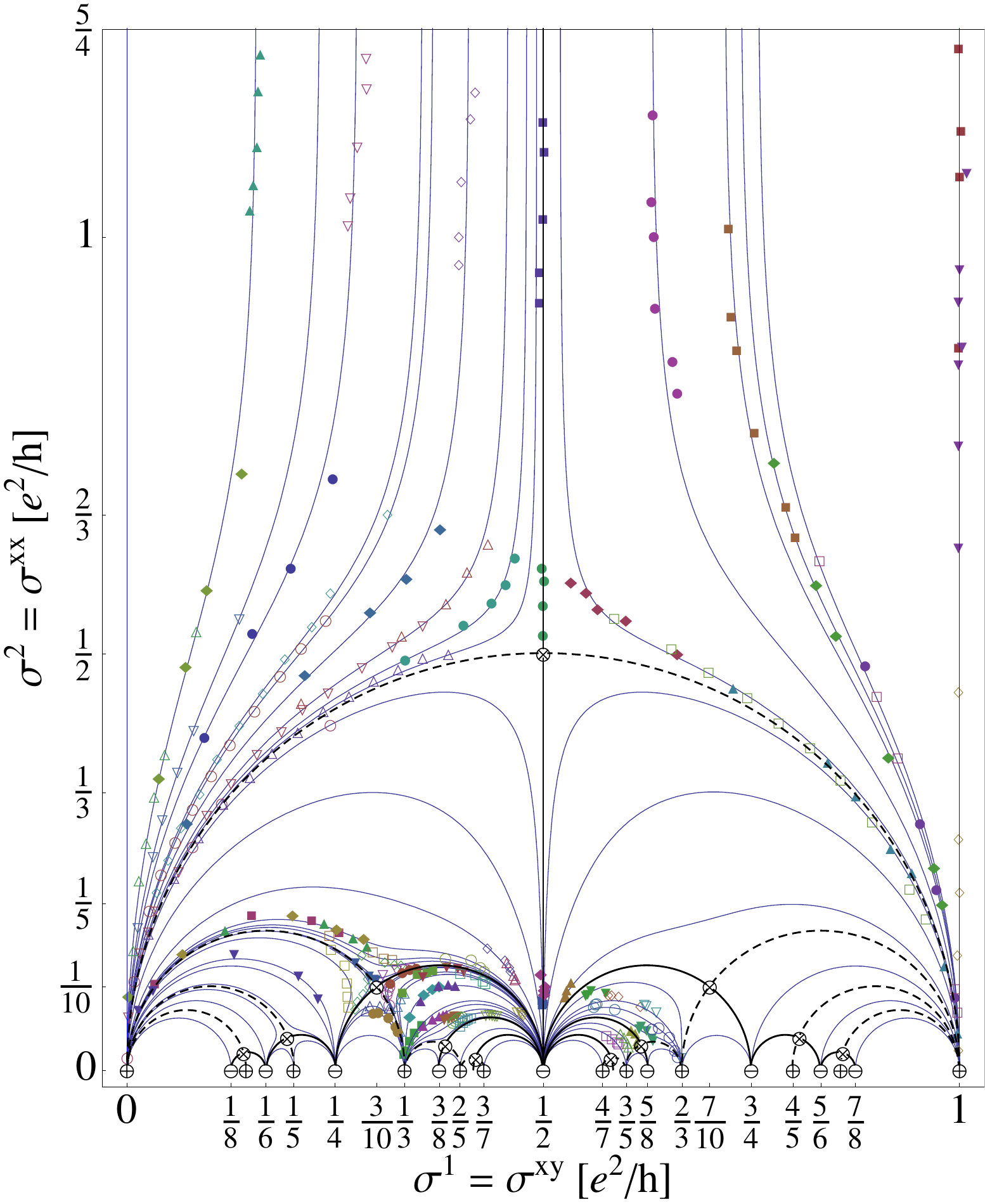}
\end{center}
\caption{Compilation of temperature-driven flow data 
\cite{Grenoble1,Grenoble2} superimposed on RG flow-lines 
derived from the RG potential $\varphi _H$. Thick black lines are phase
boundaries. Dashed lines are separatrices for the flow. The busiest region
is shown under successive magnifications in 
Fig.\ref{fig:E2flow_fig_wide} and 
Fig.\ref{fig:E2flow_fig_detail}.}
\label{fig:E2flow_fig_big}
\end{figure}

\section{Introduction}

The physics governing the plateaux transitions observed in the
quantum Hall system is still poorly understood despite more than two decades
of experimental and theoretical investigations. The transitions display a
universal scaling behaviour and the identification of their universality class
is an important part of our attempt to understand the quantum Hall system as
it should reveal what the emergent symmetries and effective degrees of
freedom of the system are, without which no deep understanding of the
physics of this system can be claimed.

In this paper we look again at the wealth of experimental data on
temperature driven flows both near and far from the critical points.
From these data
we will try to determine the global (geometrical and topological) and local
properties of the RG flow in the effective field theory (EFT) 
of a strongly interacting spin-polarized two-dimensional
electron gas.   
We will argue that these properties strongly suggests that the
quantum Hall system posesses a beautiful emergent symmetry called 
$\Gamma_{0}(2)$, and that this symmetry in turn leads to a prediction for the
effective dynamics of the system.

It has long been known that the RG fixed points of the (odd fraction and integer) quantum Hall
system are related by the global discrete symmetry $\Gamma_{0}(2)$ \cite{Oxford2}.   
This includes the stable fixed points (plateaux) where we have a reasonable understanding of the microscopic nature of the system due to work by R. Laughlin and others,  but the symmetry applies over a much greater domain of the $(\sigma^{xy},\sigma^{xx})$ conductivity plane.
This includes the quantum critical points where the subtle and ill-understood delocalisation 
of the charge carriers (presumably anyons) takes place.

Furthermore, universal scaling behaviour follows if the renormalisation group (RG) flow respects the symmetry away from the fixed points. The critical exponents obtained
from the analysis of scaling data and from numerical simulations suggests
that the RG flow is hyperbolic near critical points, i.e. the
relevant and irrelevant exponents have the same absolute value. This in turn
suggests that the symmetry acts anti-holomorphically near the fixed points 
\cite{Oxford1}. As has been emphasised before \cite{Grenoble1,Grenoble2} 
the agreement of the $\Gamma _{0}(2)$ anti-holomorphic RG flow
lines with the observed RG flow is impressive, both near and far from the
critical points, cf. Fig.\ref{fig:E2flow_fig_big}.

The geometrical form of the RG flow-lines is only part of the story, as
the rate of flow along these lines also provides important information. 
Most effort to date has
been devoted to exploring the behaviour in the neighbourhood of the fixed
points in order to determine the critical exponents. Experiments directly
determining the exponent find a universal value 
$\nu _{\mathrm{exp}} = 2.3\pm 0.1$ \cite{nu_experimental1}, and 
$\nu _{\mathrm{exp}} \approx 2.38$ \cite{nu_experimental2,Li1}. 
This is consistent with numerical studies \cite{num1,num2}.

In a previous paper we determined the geometric
exponent $\mu \approx 2.60512$ in a 
$\Gamma _{0}(2)$-symmetric toroidal model \cite{Oxford3}.
In this paper we show that by taking into account
the non-linear corrections evident in the scaling data the geometric
exponent corresponds to a delocalisation exponent 
$\nu_{\mathrm{tor}} = 2.38\pm 0.02$, which is
in excellent agreement with both the experimental and the numerical
values. This is strong evidence that the toroidal model is in the same
universality class as the quantum Hall system.

One can do more and use the toroidal model to determine the rates of flow
far from the delocalisation fixed points, and we present a comparison of these
flow rates with the available data. While the agreement is encouraging,
further experiments are needed to test the model in detail.
That the flow-lines and the flow rates should agree with the data far from
the fixed points is remarkable and requires an explanation. In our opinion
it suggests that there must be some topological structure governing the
quantum Hall system that is stable throughout the conductivity plane.

\section{Scaling}

We first introduce some notation relevant to the discussion of
renormalisation group scaling in the quantum Hall system. As mentioned above
the observed scaling behaviour is consistent with hyperbolic scaling near
the delocalisation critical points and to describe this behaviour it is
convenient to introduce complex coordinates to describe the $(\sigma^{xy},\sigma^{xx})$ conductivity plane. Choosing $z:=\sigma ^{1}+i\sigma
^{2}$ and $\bar{z}:=\sigma ^{1}-i\sigma ^{2}$, where $\sigma ^{1}:=\sigma
^{xy}$ and $\sigma ^{2}:=\sigma ^{xx}$, the corresponding differentials are 
$\partial :=\partial _{z}:=(\partial _{1}-i\partial _{2})/2$ and 
$\bar{\partial}:=(\partial _{1}+i\partial _{2})/2$, where $\partial _{i}:=\partial
/\partial \sigma ^{i}(i=1,2)$ ($\partial _{1}=\partial +\bar{\partial},
\partial _{2}=i(\partial -\bar{\partial})$). Let $\xi $ be the dominant
scale parameter that, depending on the values of the control parameters
(magnetic field, doping, etc.), could be temperature, inelastic scattering
length, the size of the sample, etc.  We have two real-valued beta-functions,
defined as usual by $\beta ^{i}:=\dot{\sigma}^{i}:=d\sigma ^{i}/dt$ $(i=1,2)$, 
where $t=\ln (\xi /\xi _{0})$, or equivalently the complex vector fields 
$\beta ^{z}:=\beta ^{1}+i\beta ^{2}=\dot{\sigma}^{1}+i\dot{\sigma}^{2}=\dot{z}$ and 
$\beta ^{\bar{z}}:=\beta ^{1}-i\beta ^{2}=\dot{\sigma}^{1}-i\dot{\sigma}^{2}=\dot{\bar{z}}$.

To discuss the general RG flow we also need local coordinates near each
fixed point: $x+iy:=z-z_{\ast }$ 
(i.e. $x(t)=\sigma ^{1}(t)-\sigma _{\ast}^{1}$ 
and $y(t)=\sigma ^{2}(t)-\sigma _{\ast }^{2}$). The beta-functions
are unchanged by this change of coordinates: $\beta ^{1}=\dot{x}=dx/dt$, 
$\beta ^{2}=\dot{y}=dy/dt$, $\beta ^{z}=\dot{x}+i\dot{y}$ and 
$\beta ^{\bar{z}}=\dot{x}-i\dot{y}$.

We start the scaling flow when $t=0$ at some initial point $z_{0} := z(0)$
in parameter space, and first study the RG-flow in the scaling domain (very
near a fixed point, i.e. $z_{0}\approx z_{\ast }$) where the flow equations
may be linearized: 
\begin{equation}
\dot{\sigma}^{i}\approx \gamma _{i}(\sigma ^{i}-\sigma _{\ast }^{i})\quad (i=1, 2).
\label{rgdelocflow}
\end{equation}
The geometric critical exponents
\begin{equation}
\mu _i:=1/\gamma _{i}  \label{rgdelocexp}
\end{equation}
parametrize the \emph{geometric} flow rates near the fixed point. Without further constraints
the flow does not enjoy any special analyticity properties: 
\begin{equation*}
z(t)=z_{\ast }+\Re (z_0-z_{\ast })e^{\gamma _1 t}+i\Im (z_0-z_{\ast})e^{\gamma _2 t},
\end{equation*}
and a more illuminating form of this equation is $y\propto x^c$ where 
$c := {\mu _1/\mu _2}$. The flow is holomorphic if $\gamma _{1}=\gamma
_{2} =: \gamma _{\ast }$, in which case we have: 
\begin{equation*}
\beta^{z}=\dot{z}\approx \gamma _{\ast }(z-z_{\ast })\implies
z(t)=z_{\ast }+(z_{0}-z_{\ast })e^{\gamma _{\ast }t}.
\end{equation*}
This may describe the flow near the UV- or IR-stable fixed points, which are
either rational, $z_{\ast }=\sigma _{\ast }^{xy}\in \mathbf{Q}$, or the
point at imaginary infinity, $z_{\infty }:=i\lim (\sigma ^{xx}\rightarrow \infty )$. 
Near these points (which compactify the upper half plane) the
topology of the phase diagram forces the flow to be vertical, so that there
indeed is only one exponent. If the exponent is negative, 
$0>\gamma _{\ast}=:\gamma _{\oplus }$ 
(positive, $0<\gamma _{\ast }=:\gamma _{\ominus }$ )
the fixed point $z_{\ast }=:z_{\oplus }$ ($z_{\ast }=:z_{\ominus }$) is
attractive (repulsive).

If we demand that the fixed point be a saddle point, which means that one
direction is attractive (contracting flow; $\gamma _{2}<0$) and one is
repulsive (expanding flow; $\gamma _{1}>0$), then the exponents have
opposite sign and the flow is given by $y\propto 1/x^{|c_{\otimes }|}$. If
the exponents also have the same absolute value ($\gamma _{\otimes }:=\gamma
_{1}=-\gamma _{2}$) the flow is \textit{hyperbolic}: $y\propto 1/x$. In this
case we shall also refer to the flow as
\textquotedblleft anti-holomorphic\textquotedblright\ because 
$\beta _{\otimes }^{z}=\dot{z}\approx \gamma _{\otimes }(\bar{z}-\bar{z}_{\otimes })$.

\subsection{$\Gamma _{0}(2)$ anti-holomorphic flow}

The group $\Gamma _{0}(2)$ acts on the complex coordinate $z$ and is
generated by a translation $T:z\rightarrow z+1$ and by the combination 
$ST^{2}S$ where $S:z\rightarrow -1/z$ is a complexified duality
transformation. As originally noted in \cite{Oxford2} the location of the
fixed points associated with the integer and odd fractional quantum Hall
states are consistent with a $\Gamma _{0}(2)$ symmetric RG flow.
Superuniversality of the scaling exponents associated with the
delocalisation fixed points follows automatically from this symmetry.

In the neighbourhood of the delocalisation fixed points the flow is found 
experimentally to be hyperbolic, i.e. anti-holomorphic \cite{Oxford1}. 
In general it need not be anti-holomorphic at higher order, but if it happens to be a 
\textit{globally} anti-holomorphic flow which also respects $\Gamma _{0}(2)$
(i.e., for which $\beta_z$ is a weight-2 automorphic
form), then it is given uniquely \cite{Oxford3} (up to a constant) by the
modular discriminant function $\Delta $: 
$\beta _{z}\propto E_{2}(z):=\partial _{z}\varphi _{H}$, 
where 
\begin{equation}
\varphi _{H}:=\frac{1}{2\pi i}\ln \frac{\Delta (2z)}{\Delta (z)}.
\label{eq:RGpotential}
\end{equation}
A practical form of $E_{2}(z)$ useful for numerical work is given by the
rapidly converging series expansion: 
\begin{equation}
E_{2}(z)=1+24\sum_{n=1}^{\infty }\frac{nq^{n}}{1+q^{n}},  \label{eq:E2}
\end{equation}
where $q:=\exp (2\pi iz)$.

In what follows we shall explore the hypothesis that this remarkable
automorphic function contains all the universal information about the
quantum Hall system, including global discrete symmetries, the existence and
location of all RG fixed points (plateaux and delocalization fixed points),
the critical exponents, the phase diagram, scaling laws, semi-circle laws,
resistivity rules, etc \cite{Oxford3}.

\section{ The geometry of the quantum Hall RG flow}

\begin{figure*}[t]
\par
\begin{center}
\includegraphics[scale = .55]{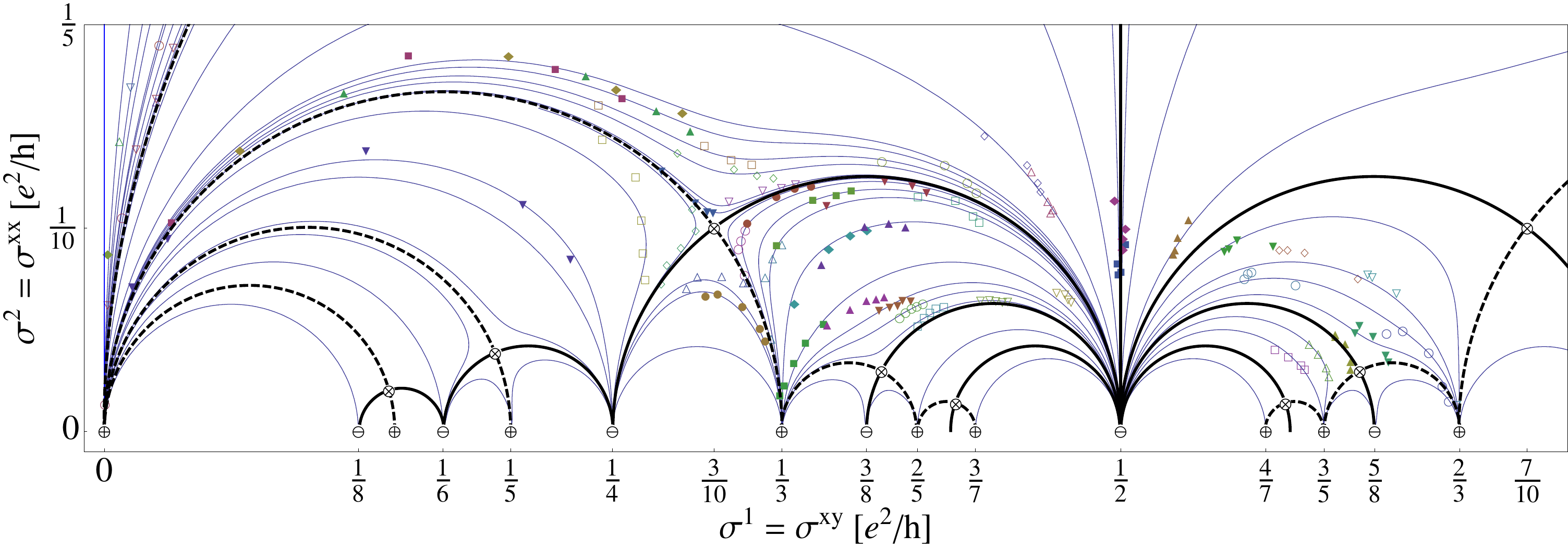}
\end{center}
\caption{Comparison of experimental temperature-driven flow \cite{Grenoble2} with our theoretical model. The data with $\sigma ^{xy}<1/2$ have been rotated by $5.5$ degrees around $(1/6,0)$ in order to align the separatrices. This systematic error may be due to a
macroscopic inhomogeneity in the sample \cite{Grenoble2}.}
\label{fig:E2flow_fig_wide}
\end{figure*}

We turn now to a comparison of the $\Gamma _{0}(2)$ holomorphic flow with
data.  We start by comparing the geometry of the predicted RG flow-lines with data,
ignoring for the moment the flow-rate. 

Fig.\ref{fig:E2flow_fig_big} shows a compilation of temperature-driven flow data 
\cite{Grenoble1,Grenoble2} superimposed on RG flow-lines derived
from the RG potential $\varphi _{H}$ defined in eq.(\ref{eq:RGpotential}).  Thick
black lines are phase boundaries. Dashed lines are separatrices for the flow
and have the characteristic form of semi-circles (\textquotedblleft the
semi-circle rule\textquotedblright ). The busiest region is shown under
successive magnifications in Fig.\ref{fig:E2flow_fig_wide} and Fig.\ref{fig:E2flow_fig_detail}. 
The integer data were obtained for temperatures
ranging over two decades, from 4.2 K down to 40 mK, for a variety of samples
and magnetic fields \cite{Grenoble1}. The diagram has been rescaled by 0.5
to remove the effect of having two flavors of charge carriers (spin up and
spin down), as explained in \cite{Grenoble1}. Data points for the flow in
fractional phases were obtained for a single sample at temperatures 
$T=0.3\mathrm{K}, 0.2\mathrm{K}, 0.11\mathrm{K}, 0.06\mathrm{K}, 0.035\mathrm{K}$ 
\cite{Grenoble2}.

Unfortunately errors are not estimated in these references. We have allowed
for some small systematic errors in both sets of data, in order to align the
experimental and theoretical separatrices. The experimental separatrix for
the integer flow is centered about 1\% below the point $(1/2,0)$, so these
data have been shifted up by $0.01$. Similarly, the experimental separatrix
for the fractional flow to the left of $\sigma ^{xy}=1/2$ appears to be
slightly rotated, so these data have been rotated by $5.5$ degrees around 
$(1/6,0)$ in order to align the separatrices. This systematic error may be
due to a macroscopic inhomogeneity in the sample \cite{Grenoble2}.

The theoretical flow-lines were obtained by numerically integrating the RG
equations: 
\begin{eqnarray}
\dot{x} &=&\Re \{Ny^{2}E_{2}(z)\}  \notag \\
\dot{y} &=&-\Im \{Ny^{2}E_{2}(z)\}  \label{rgflow}
\end{eqnarray}
where $N$ is a normalisation constant and we have used the hyperbolic metric 
\cite{Oxford3}. $E_{2}(z)$ is the holomorphic and automorphic form given in
eq.(\ref{eq:E2}), retaining 50 terms in the sum in order to get sufficient
accuracy near the real line. We get to pick the starting point for each
particular flow-line, but this completely fixes the geometry of the curve.

Given the lack of error estimates it is impossible to provide a quantitative
estimate of the goodness of fit, but qualitatively the agreement of the 
$\Gamma _{0}(2)$ holomorphic flow geometry with the data is impressive and
consistent over the range of parameter space explored so far.

\section{The RG flow rate}

Identification of the symmetry can lead to a detailed prediction of the effective
dynamics governing the flow. The observation that the symmetry acts
anti-holomorphically led us to identify a class of effective scaling models
consistent with $\Gamma _{0}(2)$, which is parametrized by the complex
structure of a torus with a special spin structure, in which only the number $f$
of flavors of fermions remains undetermined \cite{Oxford3}. Together with the
hyperbolic metric used in eq.(\ref{rgflow}) these models allow us to
determine the normalisation constant $N$ as a function of $f$.  Comparison
with the data discussed below shows that a consistent picture emerges for 
$f=2$, corresponding to  $N\approx 0.2518 i \approx i/4$, and these are the values used in the
fits presented here.

To test the resulting prediction for the RG flow rate we must go beyond the
comparison with the geometrical form of the flow-lines tested above. In
particular, it is clear from eq.(\ref{rgflow}) that the gradient of the flow-line 
is independent of the metric and normalisation, and is determined by the 
$\Gamma _{0}(2)$ function $E_{2}(z)$ only. 
Testing the flow rate is crucial to a complete test of the RG form given in 
eq.(\ref{rgflow}).

\begin{figure}[h]
\par
\begin{center}
\includegraphics[scale = .53]{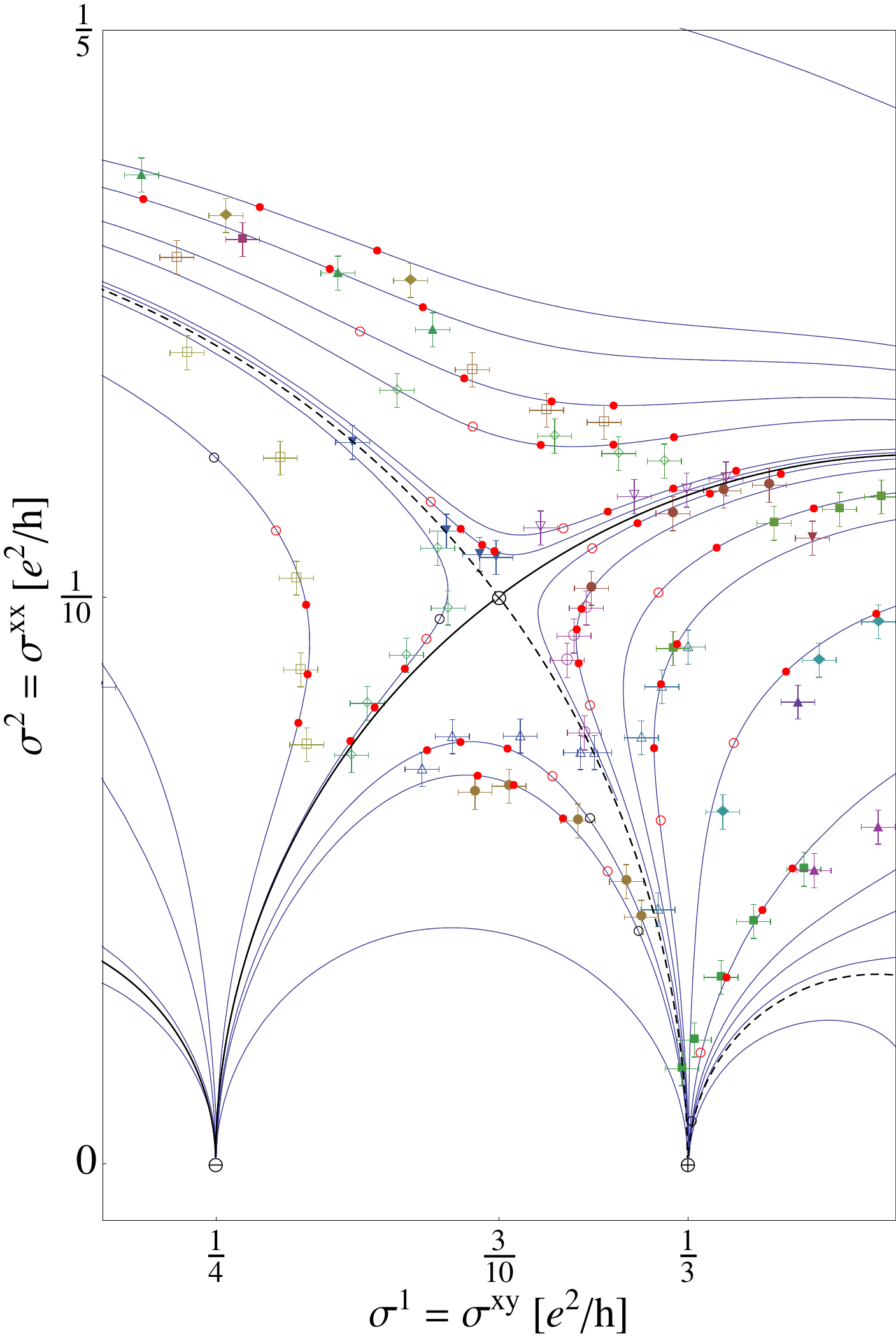}
\end{center}
\caption{Comparison of experimental temperature-driven flows 
\cite{Grenoble2} with our theoretical model. The (red/grey) bullets
and the (red/grey and black) circles are explained in the text.}
\label{fig:E2flow_fig_detail}
\end{figure}

We start with a general comparison of the predicted flow rates with the data
over the whole conductivity plane; the data in the region close to the
delocalisation fixed points is discussed in the next section. 
Fig.\ref{fig:E2flow_fig_detail} compares the experimental temperature-driven flows 
\cite{Grenoble2} with our theoretical model. We have
equipped the experimental data points in this diagram with \textquotedblleft
fake" error bars in order to distinguish them clearly from the theoretical
points derived from the toroidal model (the displayed error-bars correspond
to errors of approx. 1\% in $\sigma ^{xy}$ and 3\% in $\sigma ^{xx}$). The
data appear to follow the shape of the theoretical flow-lines quite closely.

The bullets and circles indicate the theoretical flow rate 
if the standard scaling assumptions are made in order to relate the temperatures 
$T = 0.3 \mathrm{K}, 0.2 \mathrm{K}, 0.11\mathrm{K}, 0.06 \mathrm{K}, 0.035 \mathrm{K}$ 
at which data were obtained, to the correlation lengths $\xi(T)$: 
\begin{equation}
t(T) := \ln \xi(T) = p\ln (T_0/T)/2,
\end{equation}
where $p$ is the inelastic scattering-rate exponent, and $p\approx 3.4$ is chosen
for best fit. 
As discussed below non-universal values for $p$ may be expected for the GaAs-AlGaAs 
samples studied in \cite{Grenoble2}.
The saddle point $\otimes = (3/10,1/10)$ is the quantum
critical point controlling the delocalization transition to the $\sigma^{xy}
= 1/3$ phase. The fixed points $\oplus = (1/3,0)$ and $\ominus = (1/4,0)$
are attractive and repulsive, respectively. The thick black line is a phase
boundary, while the dashed line is a separatrix of the flow.

The agreement is highly non-trivial, as it depends delicately on the choice
of metric, the scaling relations and the global value of $p$. The agreement
for the three highest temperatures (red/grey bullets:  $\sim 100$ mK) is in most cases very
good (usually within one ``fake" sigma, i.e. perfect agreement). For the two
lowest temperatures (circles: $\sim 10$ mK) the agreement is less impressive; 
the theory points are indicated by red/grey and black circles for 
$T = 0.06 \mathrm{K}, 0.035 \mathrm{K}$, respectively. 
The theory appears to \emph{systematically} underestimate the flow-rate in 
this regime, but notice that the theory is
extremely sensitive to changes in $T$. For example, if the quoted
temperatures are just $10$ mK too high the agreement is much better: almost
all points agree within two ``fake" sigmas. Notice also that the hyperbolic
metric plays an essential role. A very small change in the metric
immediately destroys the agreement. Using a flat metric is completely ruled
out.

\subsection{Critical exponents}

There have been several experiments performed in the neighbourhood of the
delocalisation fixed points to determine the critical exponents. The
starting point is the magnetic field dependence of the localisation length $\xi$ 
of electron states in the centre of an impurity-broadened Landau
Level, $\xi \propto \left\vert B-B_{\ast }\right\vert^{-\nu} $,
where $B$ is the magnetic field and $B_{\ast}$ is the critical field strength
\cite{Pruisken}.

The most direct method of measuring $\nu$ was developed by 
Koch \emph{et al.} \cite{nu_experimental1} using size dependent scaling. The experiment
studies several Hall bars with varyious widths and determines the half-width
$\Delta B$ of the $\rho _{xx}$ peak between adjacent Hall plateaux
at the point that the temperature driven flow saturates. This corresponds to
the point at which the localisation length equals the bar width and hence
the experiment directly measures the dependence of $\xi $ on $\Delta B$, from
which $\nu$ can be determined. For transitions between integer
quantum Hall states the experiment found a universal value 
$\nu_{\mathrm{exp}}=2.3\pm 0.1$.

A less direct method uses temperature dependent scaling. It measures the the
half-width $\Delta B$ of the $\rho _{xx}$ peak and the maximum slope of 
$\rho _{xy}$ as a function of temperature. These have the scaling behaviour 
$\Delta B\propto T^{\kappa }$ and $\left( \partial \rho _{xy}/\partial
(B-B_{\ast })\right) ^{\max }\propto T^{-\kappa }$, where the temperature
exponent $\kappa $ is related to the delocalisation critical exponent
by $\kappa = p/2\nu$, and $p$ is the temperature exponent of the inelastic
scattering rate. 
The half-width is defined to be the width $\Delta B$
between the inflection points of $\rho^{xx}(B)$. The empirical fact that
these points appear to coincide with the points of maximal curvature of 
$\rho^{xy}(B)$ is a consequence of the resistivity rule \cite{res-rule},
and at least in the scaling domain we may therefore expect the two graphs to
contain the same information. 

Unfortunately the measurement of the critical
exponent $\kappa$ from temperature driven flows does not give
the exponent $\nu$ directly,
since the value of $p$ must also be determined. 
For the samples studied by Koch \emph{et al.} \cite{nu_experimental1} 
non-universal values for $p$ were found ranging from $p\approx 2.7$ to $p \approx 3.4$.   
Non-universality of the exponent $\kappa$ was also found in \cite{Hohls}.

Wei \emph{et al.} \cite{nu_experimental2} 
measured $p$ in the integer quantum Hall effect by
measuring the dependence of the effective temperature of the electon gas on
the applied current.  They also obtained $\kappa \approx 0.42$ for the same sample
and hence were able to provide an independent measurement of $p$ and 
$\nu$, finding $p\approx 2$ and $\nu_{\mathrm{exp}} \approx 2.4$. 

In a very recent study Wanli Li \emph{et al.} \cite{Li1} found perfect power-law scaling with 
$\kappa = 0.42 \pm 0.01$ over two decades of temperature, and measured 
$p \approx  2$ directly from a size dependent study, corresponding to 
$\nu_{\mathrm{exp}} \approx  2.38$.

The observation of a universal value for $\kappa$ in a subset of experiments has been explained in recent work by Wanli Li \emph{et al.} \cite{Li2}. They showed that universality applies to samples with short-range disorder such as $Al_x Ga_{1-x} As/Al_{0.33} Ga_{0.67} As$ hetero-structures for $x$ between $0.65\%$ and $1.6\%$.  Universality is not found in samples in which the disorder is dominated by long-range ionized impurity potentials.

Using variable-range hopping theory Hohls \emph{et al.} \cite{Hohls} obtained  
$\nu_{\mathrm{exp}} \approx  2.35$ directly from temperature driven flows.

Temperature dependent scaling has also been studied for the transition between the fractional
quantum Hall states $2/5\rightarrow 1/3$ \cite{FQHEscaling}. A value of 
$\kappa \approx 0.42$ was also found for this case, but the value of $p$ was
not measured.

How do these measurements relate to the scaling properties of 
eq.(\ref{rgflow})? 
Close to the delocalisation fixed point the RG flow is given by 
eq.(\ref{rgdelocflow}) and the geometric exponents are defined in
eq.(\ref{rgdelocexp}). 
They are determined by the dependence of the localisation length on 
$\Delta \sigma$.  Equivalently, for the conductivity range relevant to the
Koch \emph{et al.} experiment \cite{nu_experimental1}, 
one can determine it from the same dependence of the
localisation length on the resistivity, $\Delta \rho$.  On the other hand
the delocalisation exponent is determined by the dependence of the
localisation length on $\Delta B.$ To relate the two we need to determine
the dependence of the resistivity on the magnetic field, which can be
parameterised by the exponent $\alpha $ in $\Delta \rho \propto \Delta
B^{\alpha }$.   If $\alpha =1$ then the geometric delocalization exponent
is $\nu_{\mathrm{tor}}=\mu \approx 5.210/f$ \cite{Oxford3}. 

\begin{figure}[t]
\par
\begin{center}
\includegraphics[scale = .6]{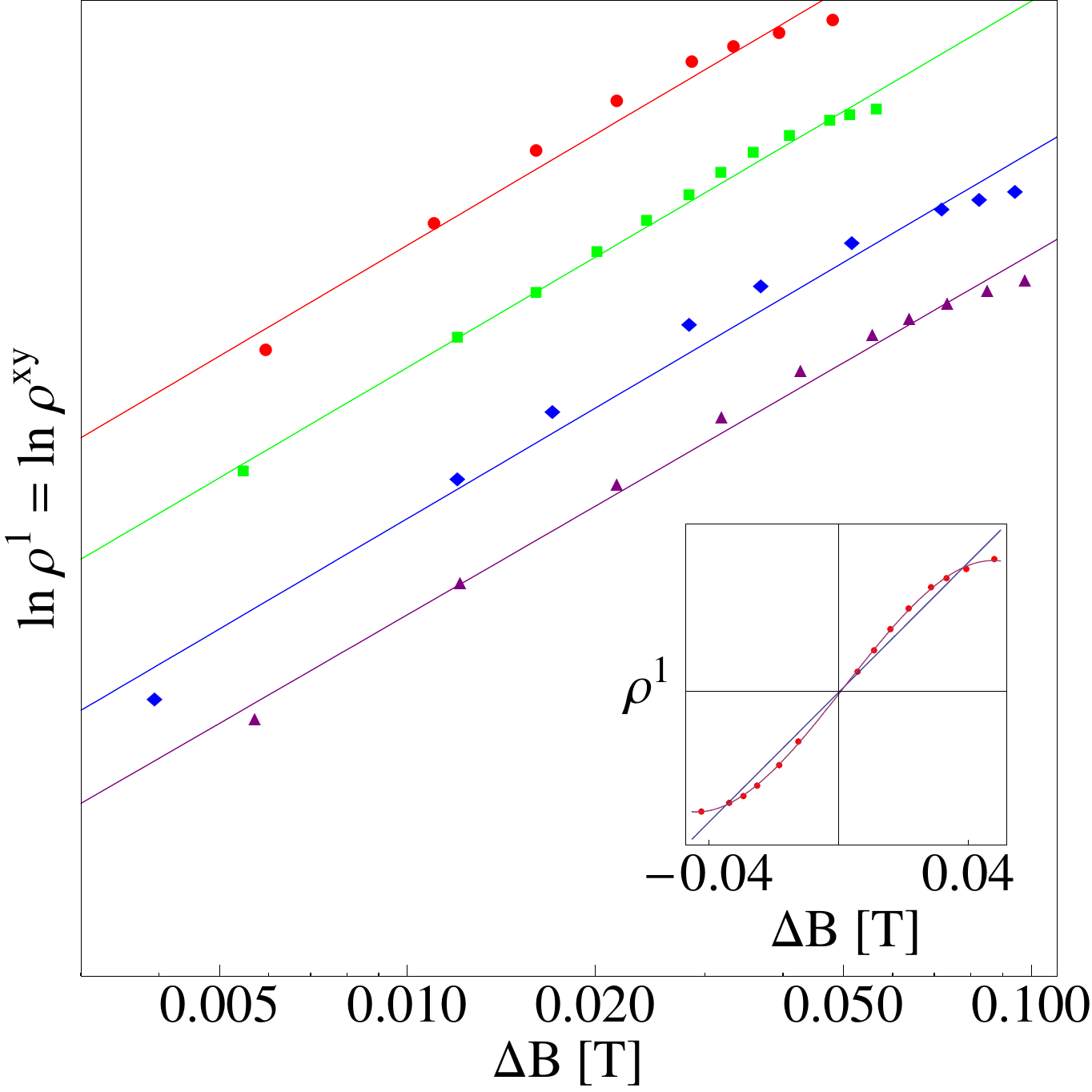}
\end{center}
\caption{Non-linear corrections to naive scaling for the four Hall-bar
geometries considered in \cite{nu_experimental1} (triangles, diamonds, squares
and bullets correspond to widths $w = 10, 18, 32$ and $64$ microns,
respectively). The inset shows  some points on $\rho^1(\Delta B) = \rho^{xy}(\Delta B)$ between the points of maximum curvature, for $w = 64\mu m$, together with chi-square fits
to a line and a cubic curve.}
\label{fig:vonk-exponents}
\end{figure}

However, as may
be seen from the inset in Fig.\ref{fig:vonk-exponents}, over the relevant
range between inflection points there is a departure from linearity. To
correct for this we determine $\alpha$ from the magneto-resistance data by
a fit (see Fig.\ref{fig:vonk-exponents}) to the experimental 
$\rho^{xy}$-curves in \cite{nu_experimental1} over the range of field values between
the points of maximum curvature, which by the resistivity rule correspond to
the inflection points of $\rho^{xx}$ that are used to define the half-width.  
The \textquotedblleft data points" in Fig.\ref{fig:vonk-exponents} are a random
but representative sample of resistivity values for $\Delta B=B-B_{\ast }>0$, 
while the straight lines are least chi-square fits. Deviations from
scaling appear far from the critical value $(\Delta B=0)$. Consistency is
verified by the fact that these lines are parallel, with a mean slope of 
$\alpha = 0.913\pm 0.008$. 

Using this gives a geometric delocalization exponent 
$\nu_{\mathrm{tor}} =  \mu \times \alpha =2.38\pm 0.02$, 
where we have used the value $\mu \approx 2.605$ for the geometric exponent 
in the bi-flavored ($f=2$) toroidal model \cite{Oxford3}. 
This is in excellent agreement with the experimental results described above 
($\nu _{\mathrm{exp}} = 2.3\pm 0.1$ \cite{nu_experimental1}, 
$2.4$ \cite{nu_experimental2},
$2.35$ \cite{Hohls},
$2.38$ \cite{Li1}), 
which appear to be converging on a value slightly below $2.4$, 
as well as the value $\nu _{\mathrm{num}} = 2.35 \pm 0.03$ \cite{num2} 
obtained from numerical simulations of this quantum phase transition.

\section{The resistivity rule and charge-flux duality}

In addition to the RG flow data discussed above there are several other
measurements that suggest a connection of the quantum Hall system with the
emergent $\Gamma _{0}(2)$ symmetry discussed here. An intriguing empirical
observation \cite{res-rule} is the \textquotedblleft resistivity
rule\textquotedblright $d\rho^{xy}/d\log B \propto n \rho^{xx}$,
where $n$ is the carrier density.
We have checked that this rule applies in the quantum regime
studied in \cite{nu_experimental1}, where $B$ (\emph{not} $\Delta B$) is
approximately constant over the transition region. It is interesting to note
that this connection between $\rho^{xy}$ and $\rho^{xx}$ follows along the 
$\Gamma _{0}(2)$ semi-circle separatrices of the RG flow connecting fixed
points if one identifies the angular variable with $\Delta B$, the applied
magnetic field difference.

Another piece of evidence pointing to an emergent $\Gamma _{0}(2)$
symmetry is the longitudinal, non-linear, current-voltage characteristics
observed near the quantum Hall liquid to insulator transition. 
It is found \cite{IV} that the $IV_{xx}$-characteristics obtained on both sides 
of some transitions (from the quantum Hall liquids with filling factors $1$ and $1/3$ 
into the neighboring insulator) map into each other under the duality transformations 
contained in the emergent $\Gamma _{0}(2)$ symmetry. 
The agreement is essentially perfect within the very small errors obtained in these 
impressive experiments (less than the pixel size in the plots), 
and probes the duality symmetry far away from the scaling domain 
close to the quantum critical points.
It will be interesting to see whether the actual shapes of the 
$IV_{xx}$-curves can be obtained directly from the emergent symmetry.

\section{Discussion}

In this paper we have reviewed the evidence coming from scaling data that
the quantum Hall system involving integer and fractional odd denominator
states has a non-Abelian discrete symmetry, $\Gamma _{0}(2)$. Numerical
studies suggest that the RG\ flow is hyperbolic near critical points
suggesting that the symmetry acts anti-holomorphically. This observation led
us to identify a class of effective scaling models consistent with this
symmetry, which is parametrized by the complex structure of a torus with a
special spin structure, in which only the number of flavors of fermions remains
undetermined \cite{Oxford3}.

Remarkably, not only do the positions of the fixed points and the RG flow-lines
agree with the data,  but the detailed flow rates also agree with the bi-flavored 
toroidal model, both close to and far away from the critical
points. Why the agreement should persist far from the critical points is not
known but strongly suggests that there is some topological order in
the quantum Hall system \cite{Oxford4}.

While this agreement with data is very encouraging, in view of the
experimental uncertainties it falls short of a definitive test of the theory based on 
the  symmetry group $\Gamma _{0}(2)$. In particular, we would dearly like more
experimental determinations of the critical exponents, particularly for the
fractional quantum Hall system where a direct measurement of the 
delocalisation exponent has yet to be performed. 
This can be done by size-dependent scaling experiments, or by
temperature dependent scaling experiments which directly measure the localisation length. 
As we have stressed repeatedly here, it is also important to test theory 
away from the critical points since this may confirm that a new topological structure
is playing a role in the quantum Hall system.

\end{document}